# Towards Providing Connectivity When and Where It Counts: An Overview of Deployable 5G Networks

Jingya Li, Xingqin Lin, Keerthi Kumar Nagalapur, Zhiqiang Qi, Adrián Lahuerta-Lavieja, Thomas Chapman,
Sam Agneessens, Henrik Sahlin, Daniel Guldbrand, Joakim Åkesson

Ericsson

Contact: {jingya.li, xingqin.lin, keerthi.kumar.nagalapur, zhiqiang.qi, adrian.lahuerta.lavieja, thomas.chapman,
sam.agneessens, henrik.sahlin, daniel.guldbrand, joakim.akesson}@ericsson.com

*Abstract*— Public safety operations require fast and reliable mission critical communications under various scenarios, in which the availability of wireless connectivity can be a question of life or death. To provide connectivity when and where it counts, we have witnessed a growing demand for deployable networks for public safety in natural disasters or emergency situations. This article investigates the opportunities of using the 5$^{th}$ generation (5G) new radio (NR) standard for designing flexible and reliable deployable networks. We describe use cases and provide an overview of deployable 5G network concepts, including architecture options, system performance analysis, and coexistence aspects. We also identify technical challenges that can be considered in the evolution of 5G NR to unlock the full potential of deployable 5G networks.

## I. INTRODUCTION

Wireless connectivity is becoming a fundamental part of our society. For public safety mission critical (MC) services, fast and reliable communications are required by first responders in various situations, in which the availability of wireless connectivity can be a question of life or death.

The 5th generation (5G) wireless access technology, known as new radio (NR), has been designed and evolved in the third-generation partnership project (3GPP) to meet a wide range of service requirements and deployments [1]. An overview of 5G NR for public safety MC communications is provided in [2], where different 5G MC services, key challenges to fulfill the MC service requirements, and different NR features being evolved to cater these challenges are discussed.

To meet the increased demands on wireless connectivity, rich features have been developed in 5G NR for further extending terrestrial mobile network coverage. Examples include using advanced multi-antenna systems together with beamforming techniques to increase signal strength in a particular direction, using repetition schemes to transmit multiple copies of the same data on different time-frequency resources or from different communication paths, using integrated access and backhaul (IAB) to provide multi-hop wireless backhaul relaying, and using sidelink for device-based relaying.

MC communications require a connectivity level that is beyond what is typically needed for consumer services. To ensure connectivity when and where it counts, there has been a growing interest in public safety communities to use deployable networks for providing temporary coverage, capacity, and redundancy in various MC situations. A deployable network is a terminology commonly used in public safety community and it refers to a wireless network that can be set up temporarily using portable/moving base stations (BSs) [3], [4]. Traditionally, deployable BSs carried on land-based vehicles, often referred to cells-on-wheels, are used to bring the temporary cellular connectivity to areas where wireless coverage was not available, permanent wireless infrastructure is not operational, or additional capacity is needed. In recent years, we have witnessed strong interest in using unmanned aerial vehicles (UAVs), such as drones, gliders, balloons, or blimps to carry network infrastructure for providing fast cellular connectivity [4]–[8]. UAV-BSs, also called as cells-on-wings, can provide more flexibility in bringing connectivity to areas that cell-on-wheels cannot reach or alleviate mobile network congestion by offloading the traffic to UAV-BSs.

In the 3GPP 4$^{th}$ generation (4G) long term evolution (LTE) standard, a feature called isolated evolved universal terrestrial access network (E-UTRAN) operation for public safety (IOPS) was introduced to maintain local connectivity for public safety users using an IOPS-capable BS when backhaul connection between the BS and the traditional core network is lost/damaged [9], [10]. The IOPS-capable BS is co-sited with or can reach a local core used in an isolated operation. The LTE IOPS solution is applicable to the establishment of deployable networks. However, the standardized solution requires a dedicated public land mobile network identity reserved for the deployable network and a separate dedicated universal subscriber identity module application configured at the device. These requirements set limitations on the configuration flexibility of deployable networks and their applicable use cases.

To unlock the full potential of deployable networks for limitless connectivity, we need to address several major challenges. For example, reliable and scalable inter-BS wireless connectivity is required when multiple deployable BSs are used to cover a wide area. Most deployable networks used in public safety rely on satellite backhaul to connect to the mobile core network, which may be costly and may not scale to large and delay-sensitive traffic demands. Some types of UAVs, like drones, have limited size, payload-weight capacity, and onboard power, which restrict the antenna design and operation time for drone-based deployable BSs. Replacement of a running deployable BS may result in service discontinuity, which is unacceptable for MC services. In addition, adding deployable BSs into an area partially covered by existing mobile networks may cause severe interference if done without careful configuration, which can impact the performance of both MC and non-MC users.

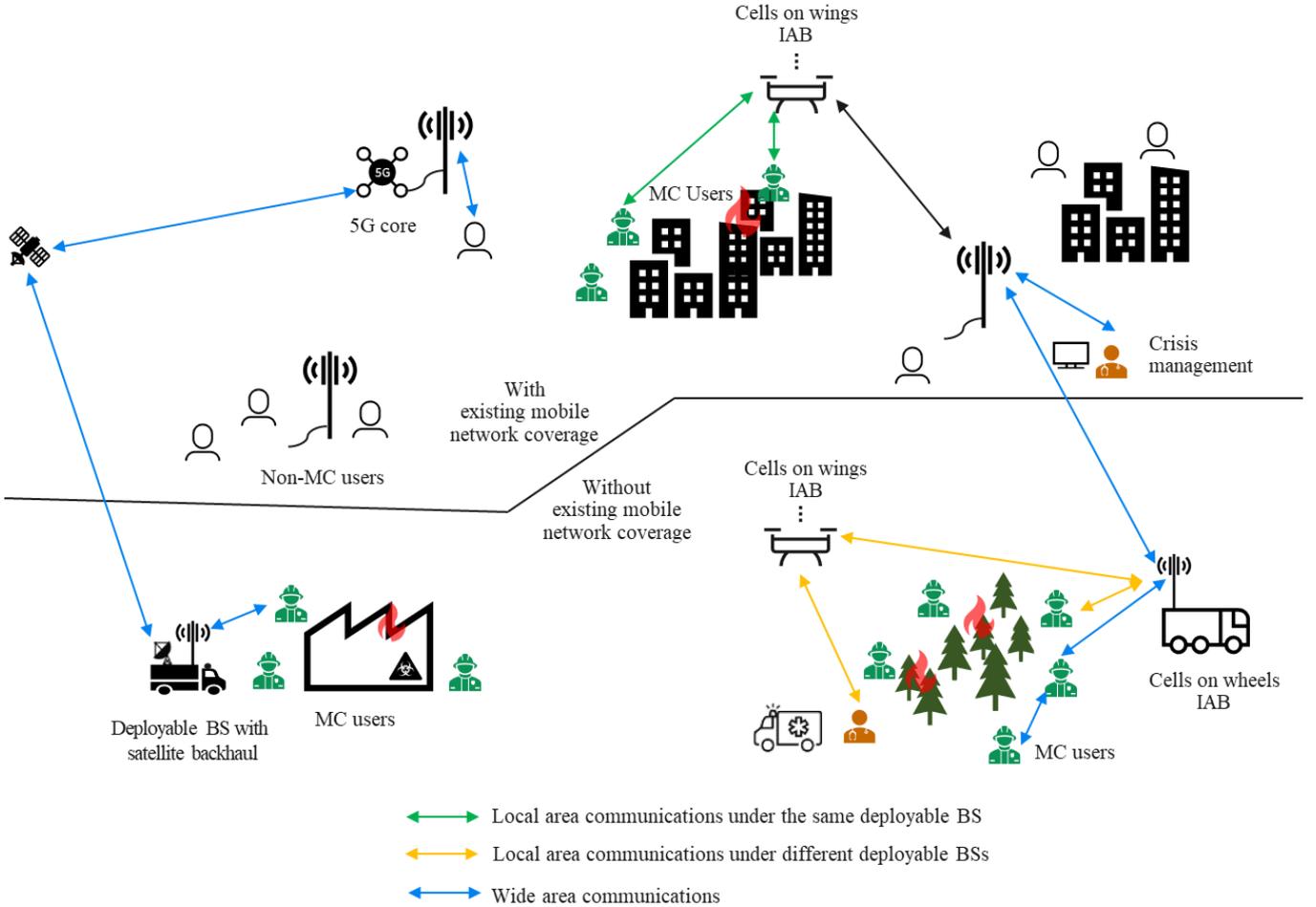

Figure 1. Illustration of basic use cases of deployable networks.

This article investigates the opportunities of using 3GPP NR features for the design of deployable 5G networks. We describe the primary use cases and give an overview of deployable 5G network concepts. Different system architecture options for establishing a deployable network are discussed and compared. System performance is evaluated to gain insights on the deployment strategy for deployable 5G networks. To improve the co-existence between deployable and existing mobile networks, different design aspects including access and admission control, antenna design for integrating UAV-based deployable BSs, and inter-network interference management are proposed. We also identify several technical challenges that require evolution of 5G NR standardization and regulation to further enhance the capabilities of deployable 5G networks.

## II. USE CASES AND ARCHITECTURE OPTIONS

In this section, we describe use cases and discuss different system architecture options for deployable networks.

### A. Use Cases of Deployable Networks

Deployable networks have been used to complement existing mobile networks to provide cellular connectivity when and where it counts. The mobile networks are generally well planned and widely deployed to provide good coverage based on a territory's population density. However, despite the widely deployed terrestrial mobile networks, providing connectivity in rural areas is challenging in many countries, mainly due to economic reasons. Deployable networks can be used in areas without existing mobile network coverage to provide temporary connectivity. Example use cases include mountain/sea search and rescue missions, forest firefighting, temporary hospitals in rural areas, disaster-struck areas where communication infrastructure is damaged, etc.

Deployable networks can also be used in areas with existing mobile network coverage, where deployable BSs can be added to boost capacity, offload traffic in mobile networks, enhance network reliability, or enhance location tracking accuracy. Example use cases include indoor firefighting where firefighters inside the building have limited connectivity to mobile networks, and big events such as festivals or concerts with high-level capacity demands.

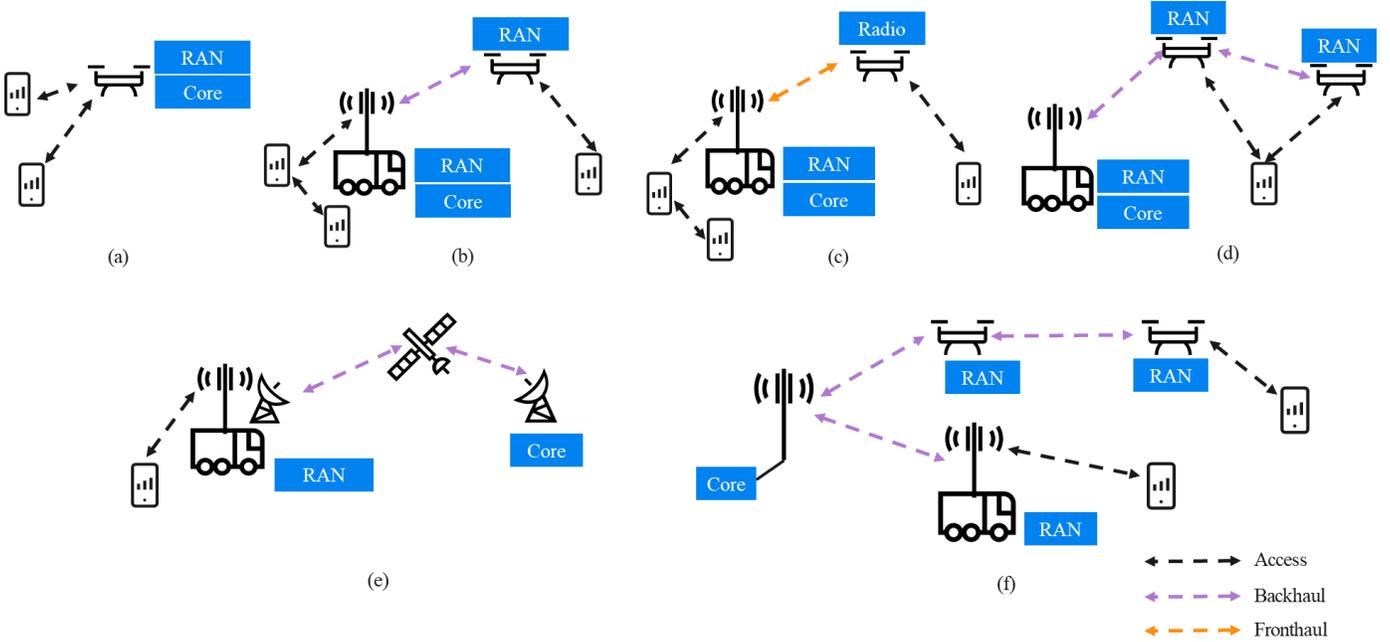

Figure 2. Illustration of architecture options for deployable networks.

As shown in Figure 1, depending on the deployment scenarios and the communication needs, deployable-network-assisted communications can be categorized into the following three basic categories:
- Local-area communications between users connected to the same deployable BS.
- Local-area communications between users connected to different deployable BSs.
- Wide-area communications between a user connected to a deployable BS and a user connected to a mobile network.

As users or deployable BSs move, deployable-aided communications may also transition within these three different categories.

*B. Standalone vs. Mobile-Network-Integrated Deployable Networks*

In general, deployable network architectures can be divided into two groups: *standalone deployable networks*, which can be operated in an isolated mode but can only provide local connectivity with limited system capacity, and *mobile-network-integrated deployable networks*, which can provide wide-area connectivity with high system capacity but requires backhaul solutions to be integrated into an existing mobile network.

*1) Standalone Deployable Networks*

Figures 2(a)-2(d) illustrate different architecture options for standalone deployable networks. Like the LTE IOPS concept and the 5G standalone architecture, a standalone deployable network has a simplified local core and an application server to host the services of interest, e.g., MC services. Due to reduced network capability and limited spectrum assets, the number of active users that a standalone deployable network can support is much smaller than what can be supported by traditional mobile networks. In addition, due to lack of connection to a mobile network, a standalone deployable network can only provide local communications between the connected users.

As illustrated in Figures 2(a), 2(b), and 2(c), for a UAV-assisted deployable network, different architecture splits can be considered: a full system on a UAV, only a radio access network (RAN) or BS on a UAV, or only the radio part on a UAV. The less hardware carried on a UAV, the longer time the UAV can fly. However, a more split architecture also implies greater communication overhead between different components and potentially higher latency.

A deployable network can consist of either a single or multiple deployable nodes. For the latter case, backhaul solutions are needed for establishing the connectivity between different deployable nodes. Compared to wired backhaul solutions, wireless backhauling can provide much more deployment flexibility. On the other hand, it is more challenging for designing wireless backhaul such that its capacity, reliability, and scalability meet the requirements of services provided by the deployable network. With 5G NR, there is an opportunity to use the IAB feature, an innovative wireless backhaul solution fully integrated with radio access, to connect different deployable nodes [11]. Compared to the 4G LTE relay solution that supports only single-hop wireless backhauling in sub-6 GHz with limited bandwidth, the 5G IAB feature supports flexible multi-hop wireless backhauling at not only sub-6 GHz but also millimeter waves carriers.

*2) Mobile-Network-Integrated Deployable Networks*

A mobile-network-integrated deployable network can temporarily extend the mobile network coverage, thereby providing wide-area connectivity to users connected to it. This is achieved by connecting the deployable BSs to the core of a mobile network using wired or wireless backhaul. As the deployable BSs are integrated into the mobile network, it is possible for the deployable network to share the mobile network

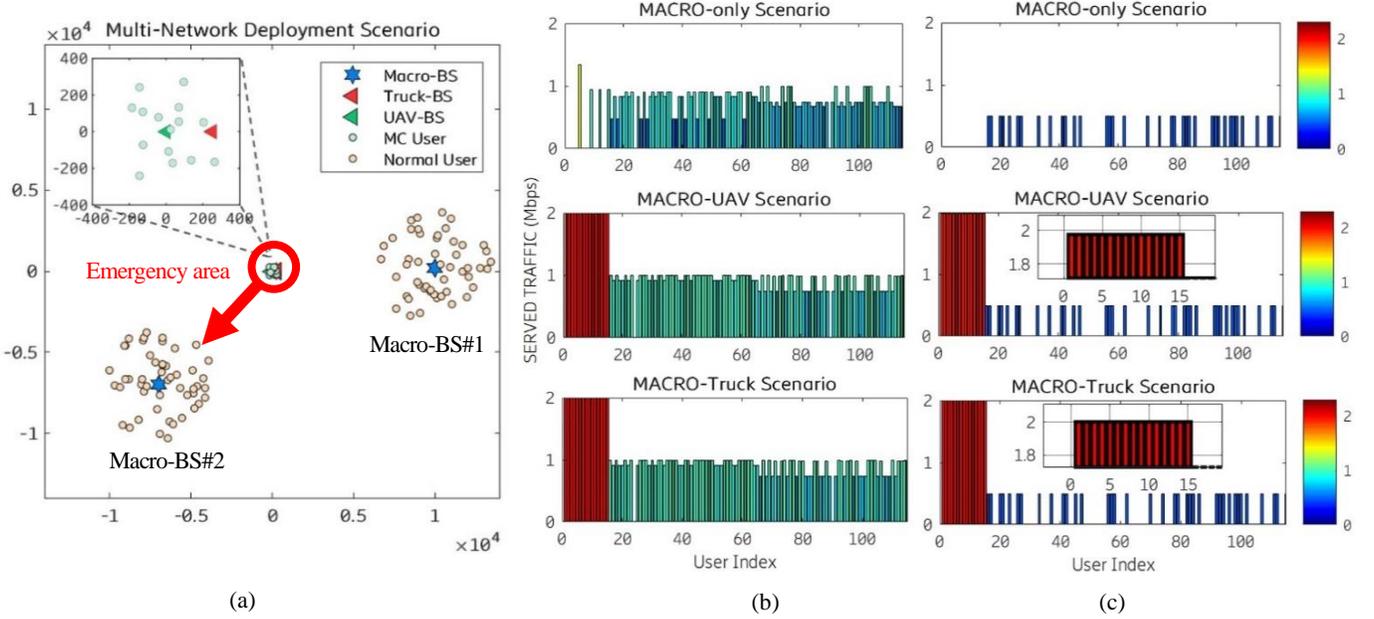

(a) (b) (c)

Figure 3. System performance evaluation results when deploying a deployable network in co-existence with a mobile network. (a) Multi-network deployment scenario [red text and shapes indicate movement of emergency area only for Figure 4]. Served (b) downlink and (c) uplink traffic per user in three scenarios.

operator's spectrum assets to provide high-capacity connectivity to otherwise unreachable users.

Most of the existing deployable networks rely on satellite backhaul to integrate the deployable BSs into a mobile network, as shown in Figure 2(e). The satellite backhaul solution is costly and does not scale well to high-capacity and time-sensitive traffic demands. To address the latency challenge, the approach of using large constellations of low earth orbit (LEO) satellites can be considered [12]. Compared to geostationary satellites, LEO satellites are deployed closer to the earth, hence, the round-trip delay can be reduced from around 600 ms to the order of a few tens of ms. However, the movement of LEO satellites leads to a varying coverage in time and space, which poses new challenges and additional complexity, e.g., for the LEO satellites to reach a ground station, for enabling UE cell selection/reselection of moving cells, and for UE mobility management.

Figure 2(f) shows another architecture alternative, where wireless backhaul connects the deployable BSs to the core network. High-capacity microwave backhaul has been the dominant wireless backhaul solution for 4G/5G mobile networks. A microwave link can provide a long backhaul range. However, setting up a reliable microwave link requires accurate measuring and tuning to align the directions at both ends of the microwave nodes, which poses constraints on the deployment flexibility and increases the time needed for network setup. The NR IAB feature is another wireless backhaul solution that can connect the deployable BSs to the core of a mobile network. This can be done by configuring a BS of the mobile network that has an existing connection to the core network as an IAB donor node and configuring the deployable BSs as IAB nodes, which are controlled wirelessly by the donor node.

Satellite and IAB are both important backhauling solutions for deployable 5G networks, and they complement each other. Satellite backhauling would be the viable solution for providing wide-area connectivity in rural areas where terrestrial backhauling does not cover, whereas for providing local-connectivity in rural areas or wide-area connectivity in urban/suburban areas with good terrestrial coverage, IAB based backhauling would outperform the satellite solution in terms of capacity, latency, deployment cost and flexibility.

### C. Flexible Network Topology Adaptation

Flexible topology is a key requirement for deployable networks to adapt to different situations and service needs in a fast and reliable way. This is especially important for public safety MC communications, where service availability and reliability need to be guaranteed under various circumstances. It is expected that new BSs can be quickly added into the deployable network for providing additional capacity, a deployable BS that is about to run out of power can be seamlessly replaced by another BS without service interruption, and in some use cases, such as a fight against a fast-moving forest wildfire, moving BSs can be used to adapt the temporary coverage area according to the needs.

The NR IAB feature is of particular interest for flexible, reliable, and reconfigurable deployment of deployable networks [2]. An IAB-based deployable BS can be self-integrated into a deployable network, utilizing the standardized IAB integration procedure. Thanks to the beamforming capabilities, there is no need for accurate measuring and tuning when setting up the wireless backhaul connection. In addition, utilizing the IAB handover procedures, seamless replacement of an IAB-based deployable BS without service interruption can be achieved, which can be particularly useful to overcome the limited operation-time constraints for drone-based deployable BSs. Moreover, to adapt to dynamic wireless backhaul link qualities and traffic load situations, and to support different system design objectives like minimizing the multi-hop latency or maximizing the system capacity, NR IAB introduced flexible topology adaptation mechanisms that allow an IAB node to change its parent node in different scenarios. This enables cells-on-wheels/wings to operate with low mobility (semi-stationary), which can be needed when the location of a deployable BS needs to be occasionally adjusted to meet the service needs.

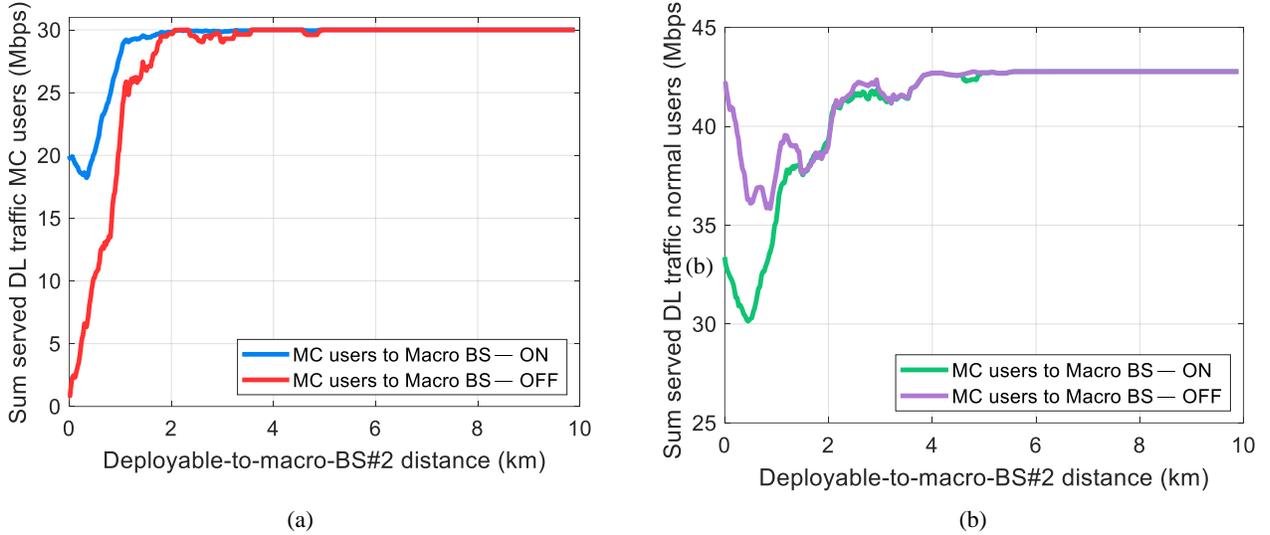

Figure 4. Impact of inter-network interference when deploying a deployable network in co-existence with a mobile network. Sum served downlink traffic for (a) MC and (b) normal users as a function of distance between the deployable BS and the macro-BS#2 (see Figure 3).

## III. SYSTEM PERFORMANCE EVALUATION AND LESSONS

In this section, we investigate the system performance when deploying a deployable network in coexistence with a mobile network. We consider a scenario shown in Figure 3(a), where a group of MC users are uniformly distributed in an emergency area located about 10 km away from the two nearest macro-BSs of a mobile network. A deployable BS (Truck-BS or UAV-BS) equipped with a local core is set up in the emergency area to provide temporary coverage to these MC users. The deployable network is operating in a standalone mode, meaning that there is no communication between deployable and mobile networks. We refer interested readers to [13].

It is assumed that a normal user can only be served by a macro-BS, while an MC user can connect to a macro-BS or a deployable BS based on the radio link quality. The required downlink (DL) traffic rates are set to 2 and 1 Mbps per MC and normal user, respectively, whereas the required uplink (UL) traffic rates per MC and normal user are set to 2 and 0.5 Mbps, respectively. The network can identify a MC user based on its access identity/category associated with its connection request, and/or the 5G quality of service (QoS) indicator associated with its MC service [14].

In a MC scenario, it is important that all the MC users are provided with the guaranteed bitrate rather than pursuing high data throughput for only a subset of users. Hence, we focus on metrics like requested traffic and served traffic for all users. A user will not be served with more traffic than required, and a user can also be dropped/blocked in case of poor link quality (quality of the wireless connectivity between a user and the network) or insufficient radio resources (e.g., due to massive connectivity request from uses).

Figure 3(b) and 3(c) show the DL and UL served traffic to all users, respectively. The x-axis of the figures represents the user indexes, where the first fifteen indexes represent the MC users, and the remaining indexes denote normal users. Three different cases are evaluated: no deployable BS (denoted by MACRO-only), a UAV-BS is deployed in the middle of the emergency area (denoted by MACRO-UAV), and a Truck-BS is deployed at the edge of the emergency area (denoted by MACRO-Truck).

As seen in Figure 3(b) and 3(c), adding a deployable BS close to MC users can significantly improve their served traffic in the considered scenarios. Compared to the MACRO-only scenario, the DL served traffic for some normal users increases after adding a UAV-BS or a Truck-BS. This is mainly because the added deployable BS can offload MC traffic from the macro-BSs, thereby freeing up resources of the macro-BSs for serving normal users. Moreover, as shown in the zoomed-in part of each sub-plot in Figure 3(c), for the MACRO-Truck scenario, all MC users can be served with the required UL traffic, while the served UL traffic for MC users are slightly below 2 Mbps for the MACRO-UAV scenario. This is mainly due to different capabilities of UAV-BS and Truck-BS assumed in the simulation setting, i.e., different transmit power levels (10 W for Truck-BS and 0.25 W for UAV-BS) and different number of BS sectors (3 for Truck-BS and 1 for UAV-BS).

To observe the impact of network interference on MC traffic performance, we focus on the MACRO-UAV scenario and vary the distance between one of the macro-BSs and the emergency area, by moving the emergency area towards macro-BS#2, see Figure 3(a). We consider two cases, i.e., to allow and not to allow the MC users to access the existing mobile network. Figure 4(a) and 4(b) show the sum of DL served traffic to MC and normal users as a function of the distance between macro-BS#2 and the UAV-BS, respectively. We see that when the UAV-BS is deployed far away from the macro-BSs, i.e., from 5 km and further, the inter-network interference is negligible. However, when the UAV-BS is located close to a macro-BS, the inter-network interference can result in service disruption of both normal and MC users. For normal users, as the UAV-BS moving closer to the macro-BS, the served DL normal traffic firstly decreases due to increased inter-network interference, then, it increases since fewer MC users are being served by the UAV-BS due to outage caused by interference.

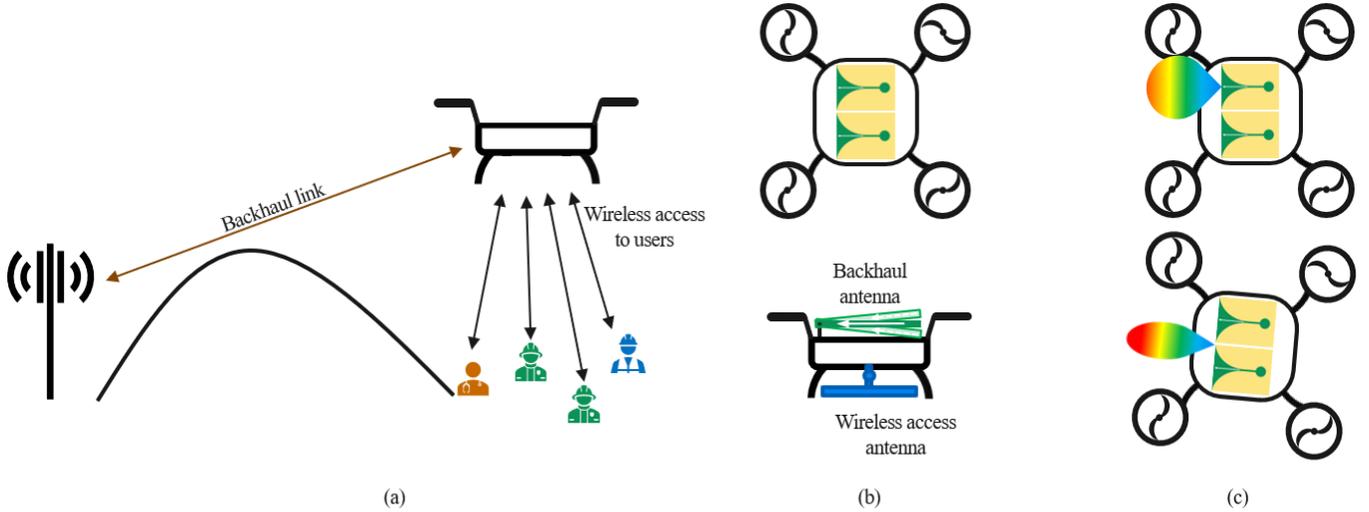

Figure 5. An antenna design for UAV based Deployable BSs favoring both backhaul and access links.

From Figure 4(a), we also observe that it is desirable to allow MC users to connect to the existing mobile network to benefit from potentially better link connections to the macro-BSs in certain deployments. For this, multi-network interference mitigation and prioritized MC traffic handling, discussed in more detail in the next section, become critical aspects to ensure the performance of MC users.

## IV. CO-EXISTENCE WITH EXISTING MOBILE NETWORKS

In this section, we discuss important design aspects that should be considered in scenarios where deployable networks are in co-existence with existing mobile networks.

### A. Access and Admission Control

Access and admission control are essential to preserve a stable working condition of a deployable network and to ensure service quality of the critical users.

A standalone deployable network can only provide local connectivity between users that are connected to it. 5G access control mechanisms like cell reservation and unified access control can be used to ensure that the deployable network is only accessible for a group of interested users. Like the LTE IOPS solution, different cell reservation indicators broadcasted in the system information message can be configured to allow only operator's users or an authorized group of users to access the deployable network.

A mobile-network-integrated deployable network can provide wide-area connectivity. Hence, it is beneficial to allow normal users to connect to it as long as it does not affect critical users accessing the network and the QoS for ongoing critical traffic can be maintained. To achieve this, leveraging different user/service-type information and priority indicators from the 5G standard, such as access identity, access category, establishment cause, 5G QoS indicator, allocation and retention priority, and network slice indicator, a mobile-network-integrated deployable network can carefully apply different admission control tools like prioritized random access, differentiated scheduling, load balancing, or resource reservation and partition to prioritize the critical users/services and preempt non-critical traffic when needed [14].

### B. Antenna Design for UAV-BSs

The antenna system on a UAV-BS plays a crucial role when integrating it into a deployable network, since it directly influences the quality of both the wireless access and backhaul links, as shown in Figure 5(a). When a wireless backhaul is used to provide core connectivity, the gain of the antenna system used for the backhaul link determines the maximum possible bitrate and the distance at which the UAV-BS can be flown. Therefore, it is beneficial to have large antenna gain, which in turn translates to a large form factor since the antenna gain is proportional to the size for a given carrier frequency.

Antenna design for UAV-BS is greatly influenced by the form factor and carrying capacity of the UAV in question [15]. Antenna design for UAV-BS based on high altitude platforms and blimps has been well explored in both literature and several prototypes. The ability of these forms of UAV to carry large payloads and the higher altitudes provide flexibility to mount all the necessary antennas on the belly of the vehicle without a tight constraint on volume and weight. However, a deployable BS based on a light and compact UAV poses tighter payload weight and volume constraints. Furthermore, the antenna system needs to be designed such that the balance and the maneuverability of the UAV are not compromised.

To achieve large antenna gains for the backhaul link on light UAV-BSs, end-fire antennas with carefully chosen mounting position on the UAV-BSs are an attractive option. The radiation pattern of such antennas has its main lobe along the length of the antenna, as opposed to the broad-side designs where the main beam is perpendicular to the plane of maximum dimension. Printed Yagi-Uda or Vivaldi antenna designs are two examples that exhibit the desired end-fire property in addition to being lightweight. Exploiting these properties, a single element or an array of end-fire antennas can be mounted above or below the light UAV-BS, as shown in Figure 5(b). In addition to being light and offering large antenna gain, the design offers benefits in terms of low wind resistance compared to an antenna system with a large aperture perpendicular to the body of the UAV. Such a design is especially beneficial at lower carrier frequencies, e.g., public safety band 14 (~700 MHz), where a broadside antenna system with equivalent gain would have a large aperture and would have to be mounted on the side of the UAV. The ability of a UAV to rotate in the azimuth direction can be used to align the main beam of the antenna towards the donor node. This keeps the complexity and cost of the antenna system low since there is no need for a beam steering network. The alignment procedure of the beam can be performed in a stepwise manner when multiple elements are employed, as shown in Figure 5(c). In step 1, a subset of elements is switched on for a broader beam, and the UAV is oriented to align with the donor node. In step 2, all the elements are activated to achieve a high gain, narrow beam, and the UAV is oriented to align the narrow beam with the donor node. In case the UAV cannot maintain precise orientation towards the donor node, e.g., in the case of strong winds, the UAV-BS can revert to the broader beam to provide resilience while sacrificing antenna gain.

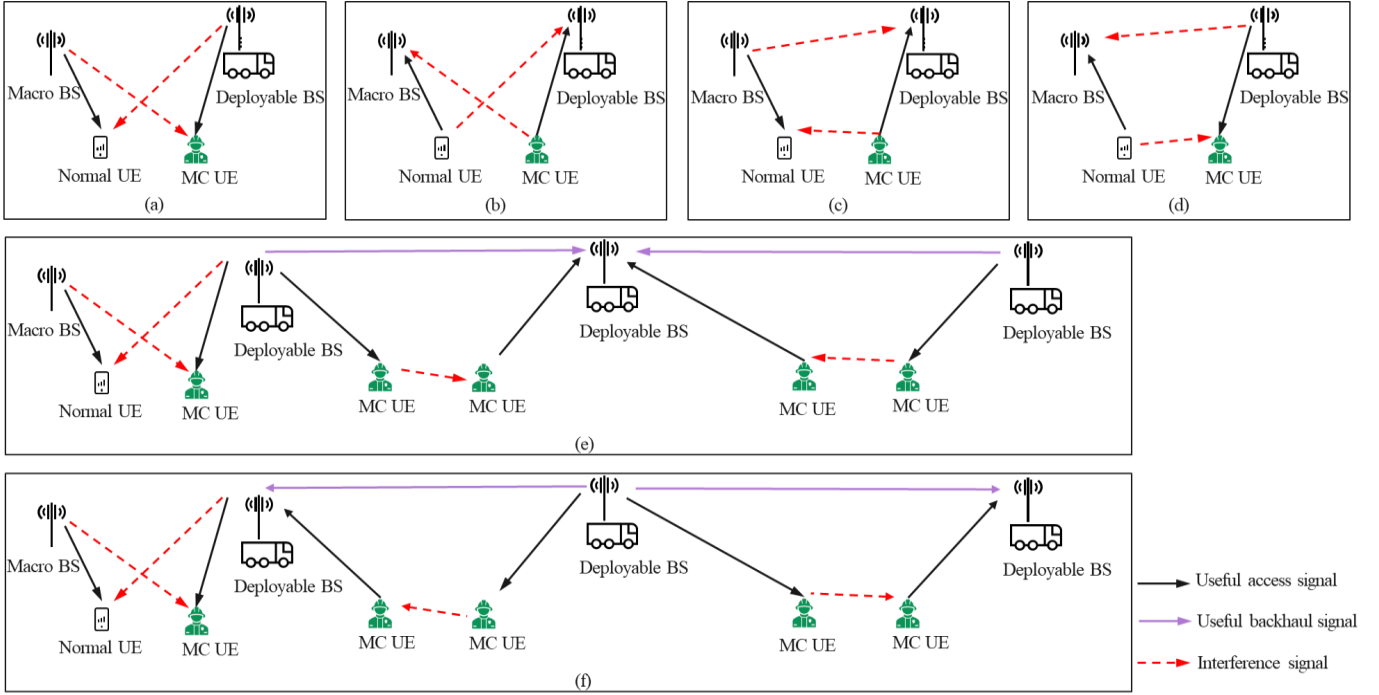

Figure 6. Different interference scenarios when a deployable BS(s) is placed close to an existing mobile macro-BS. (a) and (b) Inter-network interference when the deployable and macro-BSs are synchronized (operating in the same direction, i.e., UL or DL). (c) and (d) Cross link interference (inter-user interference and inter-BS interference) when the deployable and macro-BSs are unsynchronized (one operating in UL and the other operating in DL). (e) and (f) Cross link interference when multiple deployable BSs are wirelessly connected using IAB.

### C. Inter-Network Interference Management

As shown in Section IV, when adding deployable BSs into an area partially covered by existing mobile networks, it is essential to manage the interference between the two networks to ensure that it will not cause service degradation, especially for MC users.

In case both deployable and mobile networks are operating in frequency division duplex (FDD), the interference situations are similar to traditional FDD inter-cell interference situations (e.g., HetNet use cases), as illustrated in Figure 6(a) and 6(b). If at least one of the networks is operating in time division duplex (TDD), or both networks are operating in TDD and they are unsynchronized, besides the interference situation described for the FDD case, cross link interference (CLI) between the two networks will be present (see Figure 6(c) and 6(d)). CLI may also occur in scenarios where IAB is used to support multi-hop wireless backhauling between different deployable BSs (see Figure 6(e) and 6(f)). Like dynamic TDD use cases, two kinds of CLI can arise: inter-user interference and inter-BS interference. While it is obvious that these types of interference can happen when the deployable and mobile networks are operating on the same frequency, they can also occur and cause performance degradation when two networks are on adjacent frequency.

For a mobile-network-integrated deployable network, as the deployable BSs are integrated into the mobile network, it is easier to achieve synchronization between the two networks and avoid/coordinate simultaneous transmission and reception by different BSs. Most of the traditional multi-cell interference management schemes can be reused. Examples include coordinated multi-point transmission and reception, coordinated scheduling and power control, radio resource partition, advanced interference rejection combining receiver, etc. For the CLI case, interference measurements from a specific set of BSs/users can be used to assist a BS or a network in understanding the CLI situation.

Interference coordination between standalone deployable networks and mobile networks is more challenging since there is no communication between the deployable and mobile networks. Without knowing when and where the standalone deployable network will be deployed and how it will be configured, it can be difficult for the mobile network to properly configure its users to perform inter-network interference measurements or to perform the measurements by itself.

## V. Standardization Aspects

In this section, we summarize design aspects that require evolving 5G standardization to unlock the full potential of deployable networks.

**IAB enhancements**: To better support deployable networks with mobile deployable BSs, the IAB feature can be further enhanced to support mobile IAB operations. Using mobile IAB will, however, result in increased complexity of IAB-based deployable solutions, since handling user connection, moving interference, and service interruption will become more challenging in dynamic network topologies. Another interesting and somewhat contradicting direction for IAB enhancements is to consider reduced complexity/capability for portable IAB solutions so that it is possible to use an IAB-based deployable BS for cells-on-wings operations in practice.

**Access control enhancements**: As a standalone deployable network typically has a much lower capability compared to a traditional cellular mobile network, there can be situations where the system capacity provided by the deployable network is not capable to serve all interested users/services. In such cases, 5G access control mechanisms need to be enhanced to allow finer priority differentiation between different interested users and their requested services. Then, a standalone deployable network can wisely select which of the prioritized users/services shall access the system at an early stage.

**Multi-network interference handling enhancements**: A group of authorized users can be allowed to connect to both a standalone deployable network and a mobile network. In this case, 5G standards can be enhanced to enable an authorized user to measure and report the inter-network interference, which can be used for assisting the two networks to detect the interference and make proper interference management decisions. The regulators may also authorize an interested user to mandate one of the networks to take interference mitigating actions.

## VI. Conclusions and Future Work

Using deployable cellular networks to provide temporary and on-demand connectivity has recently gained momentum in both academic and public safety communities. The inherent flexibility of 5G NR offers a unique opportunity to develop reliable and flexible deployable networks for providing connectivity when and where it counts. In this article, we provide a detailed technical overview of deployable 5G network concepts. To further enhance the capabilities of deployable 5G networks, there are several aspects including flexible wireless network topology adaptation, early access control with finer priority differentiation, and inter-network interference handling that can be considered in further research and standardization development.

BIOGRAPHIES

**Jingya Li** (jingya.li@ericsson.com) is a Master Researcher at Ericsson Research, Gothenburg, Sweden. She is currently working with 5G/6G research and standardization in the areas of AI/ML for PHY and public safety communications. Jingya has led different back-office feature teams for 5G New Radio (NR) standardization. Her contribution to standardization has been on 5G NR for AI on PHY, public safety, initial access, remote interference management and cross link interference handling, vehicle-to-anything communications, and latency reduction. She is co-author of the book "5G and Beyond: Fundamentals and Standards." She received the IEEE 2015 ICC Best Paper Award and IEEE 2017 Sweden VT-COM-IT Joint Chapter Best Student Journal Paper Award. She holds a Ph.D. degree (2015) in Electrical Engineering from Chalmers University of Technology, Gothenburg, Sweden.

**Xingqin Lin** (xingqin.lin@ericsson.com) was a Master Researcher at Ericsson and a member of the Ericsson NextGen Advisory Board. He is co-author of the book "Wireless Communications and Networking for Unmanned Aerial Vehicles" and the lead editor of the book "5G and Beyond: Fundamentals and Standards." He received the 2020 IEEE Communications Society Best Young Professional Award in Industry, the 2021 IEEE Vehicular Technology Society Early Career Award, and the 2021 IEEE Communications Society Fred W. Ellersick Prize, among others. He is an IET Fellow. He holds a Ph.D. in electrical and computer engineering from The University of Texas at Austin.

**Keerthi Kumar Nagalapur** (keerthi.kumar.nagalapur@ericsson.com ) is a Senior Researcher at Ericsson Research, Gothenburg, Sweden. His current research activities focus on 5G NR for public safety communications and high-speed train communications, and multi-antenna enhancements in 5G NR. He holds a Ph.D degree in Electrical Engineering from Chalmers University of Technology, Gothenburg, Sweden.

**Zhiqiang Qi** (zhiqiang.qi@ericsson.com) is an experienced researcher at Ericsson Research, Beijing, China. His current research activities focus on 3GPP standardization of UAV work item, and performance evaluation of vehicle network mainly in the aera of public safety, resource allocation and interference mitigation. He received the B.S., M.S., and Ph.D. degrees from Beijing University of Posts and Telecommunications, China.

**Adrián Lahuerta-Lavieja** (adrian.lahuerta.lavieja@ericsson.com) is an Experienced Researcher at Ericsson Research, Gothenburg, Sweden. His current research interests include public safety and mission critical communications, deterministic channel modeling, artificial intelligence, and aerial communications. He received the M.Sc. degree in telecommunication engineering from the University of Zaragoza, Zaragoza, Spain, in 2016 and the Ph.D. degree in electrical engineering from KU Leuven, Leuven, Belgium, in 2021.

**Thomas Chapman** (thomas.chapman@ericsson.com) is a Principle Researcher within the radio access and standardization team within the Standards and Technology group at Ericsson. He has been contributing into 3GPP standardization since 2000 to the whole portfolio of 3GPP technologies including UTRA TDD, WCDMA, HSPA, LTE and NR, and has been deeply involved in concept evaluation and standardization of AAS and NR in RAN4. He holds an MSc (1996) and PhD (2000) in electronic engineering and signal processing from the University of Manchester, UK.

**Sam Agneessens** (sam.agneessens@ericsson.com) received the M.S. and PhD degrees in electrical engineering from Ghent University, Belgium, in 2011 and 2015. He was a postdoctoral fellow of Research Foundation-Flanders (FWO-V) affiliated with Ghent University and imec and became assistant professor at Ghent university in 2017. He joined Ericsson AB (Gothenburg, Sweden) in 2018 as Senior Researcher.  His research background is in electromagnetics focusing on robust antenna systems for wearable applications and mmWave systems. He currently leads the wireless transport research project at Ericsson. Dr. Agneessens received the URSI Young Scientist Award at the 2014 URSI General Assembly and was awarded the 2014 Premium Award for Best Paper in IET Electronics Letters.

**Henrik Sahlin** (henrik.sahlin@ericsson.com) is research manager at Ericsson Research, Gothenburg, Sweden, with focus on multi-antenna solutions, Over-the-Air measurements, backhaul and propagation. Previously he was project manager with focus on automotive, transport and public safety related research. He has been actively participated in 3GPP (Third Generation Partnership Project) standardization for LTE (Long Term Evolution) and NR (New Radio, 5G) with a focus on initial access and reduced latency with physical layer design. Receiver algorithms in base stations and mobile devices has been a central research area for Henrik within several wireless standards such as NR, LTE, WCDMA and GSM. Henrik holds a Ph.D. within the area of signal processing in electrical engineering from Chalmers University of Technology in Gothenburg, Sweden.

**Daniel Guldbrand** (daniel.guldbrand@ericsson.com) is a portfolio manager for Mission Critical Networks at Ericsson. He has over 25 years of experience from research and development within telecommunication with focus on products and solutions targeting mission critical operations and users. Daniel has studied electrical engineering at the university of Örebro.

**Joakim Åkesson** (joakim.akesson@ericsson.com) is a Senior Specialist in mission critical networks at Ericsson. Since joining Ericsson in 2008 he has been instrumental in providing solutions to Public Safety and other Mission Critical segments facilitating the adoption of 4G and 5G based networks for mission critical communication. He holds a M.Sc. degree (1995) in Applied Physics & Electrical Engineering from Linköping University of Technology.